\documentclass{article}

\usepackage[nonatbib,preprint]{neurips_2025_modified} 
\usepackage[numbers,sort&compress]{natbib}

\usepackage[utf8]{inputenc} 
\usepackage[T1]{fontenc}    
\usepackage{hyperref}       
\usepackage{url}            
\usepackage{booktabs}       
\usepackage{amsfonts}       
\usepackage{nicefrac}       
\usepackage{microtype}      

\usepackage{subcaption}%
\usepackage{graphicx}
\usepackage{colortbl}
\usepackage{makecell}

\usepackage[usenames,dvipsnames]{xcolor}
\usepackage{enumitem}
\hypersetup{ %
	pdfauthor={},
	pdftitle={},
	pdfsubject={},
    pdflang={en},
	bookmarks,
	bookmarksnumbered,
	backref=page,
	pageanchor=true,
	plainpages=false,
	colorlinks=true,%
	linktocpage=true,
	citecolor=MidnightBlue,%
	filecolor=BrickRed,%
	linkcolor=ForestGreen,%
	urlcolor=ForestGreen %
}

\newcommand{\texttalktitle}[1]{{\emph{#1}}}

\makeatletter

\newcommand{\Rmnum}[1]{\expandafter\@slowromancap\romannumeral #1@}
\makeatother

\title{Human-AI Co-Creativity:\\Advances, Opportunities, and Challenges\\
		\vspace{4mm}
		\normalsize{\textnormal{Overview of the workshop activities at ICML 2026 conference\thanks{\url{https://genaicreativity.org/icml2026/}; Corresponding author: <adishs@mpi-sws.org>.}}}
}

\author{
  \textbf{Adish Singla}\textsuperscript{1} \ \ \ 
  \textbf{Abhilasha Ravichander}\textsuperscript{1} \ \ \  
  \textbf{Liwei Jiang}\textsuperscript{2} \ \ \ 
  \textbf{Alexander Spangher}\textsuperscript{3} \ \ \ 
  \textbf{Alice Oh}\textsuperscript{4} \\
  \textbf{Jiho Jin}\textsuperscript{4}\thanks{Student volunteers as part of the workshop organization team.} \ \ \ 
  \textbf{Jun Seong Kim}\textsuperscript{4}\footnotemark[2] \ \ \ 
  \textbf{Changyoon Lee}\textsuperscript{4}\footnotemark[2] \ \ \ 
  \textbf{Manh Hung Nguyen}\textsuperscript{1}\footnotemark[2] \ \ \ 
  \textbf{Chao Wen}\textsuperscript{1}\footnotemark[2]\\  
  \\
  \textsuperscript{1} Max Planck Institute for Software Systems, Germany \\
  \textsuperscript{2} University of Washington, USA \\
  \textsuperscript{3} Stanford University, USA \\
  \textsuperscript{4} KAIST, South Korea \\
}

\begin{document}

\maketitle

\begin{abstract}
This survey article has grown out of the human-AI co-creativity workshop organized by the authors at the ICML 2026 conference. We organized this workshop as part of a community-building effort to bring together researchers and practitioners interested in topics of generative AI, creativity, and human-AI co-creation. This article aims to provide an overview of the workshop activities and highlight several future research directions in the area of human-AI co-creativity.
\end{abstract}


%

\section{Introduction}
\label{sec.introduction}

Recent advances in large generative models have turned AI agents into everyday companions. Millions of people now rely on them across domains from design and communication to science and education. 
On the one hand, these advances offer unprecedented opportunities to support people in open-ended domains by providing a medium for brainstorming ideas, exploring design choices, and thereby improving their creative outcomes.
On the other hand, existing models and interfaces are primarily designed for automation, and their increasing usage for creative tasks poses new challenges, such as the risk of design fixation and idea homogeneity. More concerningly, the integration of generative AI agents into creative workflows further raises serious issues of content copyright and authorship.

For us to fully realize these opportunities and tackle these challenges, it is crucial to build a community comprising researchers, industry professionals, and practitioners that are ``multilingual'' with (a) technical expertise in the cutting-edge advances in generative AI, (b) know-how of designing/deploying generative AI systems for human users, and (c) first-hand experience of working with these systems for creative content creation in everyday work.

The goal of the workshop has been to foster such a multilingual community by bringing together researchers and practitioners interested in topics of generative AI, creativity, and human-AI co-creation.  More concretely, the workshop brought together speakers and participants with diverse backgrounds: (i) researchers in machine learning, generative AI, natural language processing, human-computer interaction, industrial design, social robotics, and human psychology; (ii) industry professionals working on end-user systems such as Midjourney; and (iii) practitioners leveraging generative AI in art and creative industries. 
The workshop investigated these opportunities and challenges by focusing discussions along two thrusts:
%
%
\begin{enumerate} [label={\Roman*.}, parsep=4pt, leftmargin=*,labelindent=-6pt]
\item \looseness-1Exploring how advances in generative AI provide new opportunities to support people in open-ended creative tasks to brainstorm ideas and create content.
\item Identifying unique challenges in integrating generative AI into creative workflows, including design fixation, idea homogeneity, and issues of content copyright and authorship.
\end{enumerate}

By fostering collaboration between different communities and stakeholders, the workshop sought to facilitate the development of next-generation technologies that enhance human-AI co-creativity.

%

\section{Overview of the Workshop Activities}
\label{sec.workshop_summary}

\looseness-1In this section, we provide an overview of the workshop organized at the ICML'26 conference; full details are available on the \href{https://genaicreativity.org/icml2026/}{workshop website}. 
The workshop bridged ideas between generative AI (a central topic in machine learning) and creativity (integral to many real-world domains). Given that both topics are highly timely and of great significance to the ICML community, the workshop attracted considerable interest and engagement. The workshop program, including a panel and poster session, and the physical environment at the ICML'26 venue in Seoul provided ample opportunities for extensive discussions and networking among attendees.

\subsection{Topics of Interest}
As mentioned above, the workshop focused on two thrusts,  each covering several topics of interest. These topics served as guidelines when selecting the speakers for invited talks and when accepting the contributed papers.

The topics in the first thrust focused on leveraging recent advances in generative AI to support people in open-ended creative tasks, including (i) sharing viewpoints, artifacts, or field experiences of using generative AI in creative domains; (ii) exploring the creativity and diversity of generative AI models across different domains, including everyday open-ended tasks, art creation, game development, or scientific ideation; (iii) developing novel methods to improve the creativity/diversity of generative models, e.g., fine-tuning using creativity-centric rewards and inference-time methods using persona conditioning; (iv) designing new architectures for creativity-centric generative models, in particular, that are more suitable for human-AI co-creation, e.g., by leveraging insights from literature on human creativity; (v) curating benchmarks and shared resources that focus on evaluating the creativity aspects of AI agents, both when acting alone and when co-creating with a human user in the loop.

The topics in the second thrust focused on unique challenges when integrating generative AI agents into creative workflows, including (i) sharing viewpoints or field experiences about concerns in using generative AI in creative domains, and longer-term consequences of generative AI on the future of creative professions; (ii) exploring novel interfaces and interaction paradigms that tackle the challenge of design fixation or idea homogeneity, and boost creativity in human-AI co-creation; (iii) developing novel safeguarding methods to validate the authenticity of content, e.g., to determine whether a news story was written by a human or generated by a model; (iv) investigating how to attribute and reward human creativity when co-creating with generative AI.

\subsection{Accepted Papers and Poster Session}

In our call for papers and participation, we invited submissions of various types, including research papers reporting new results, position papers presenting novel viewpoints or field experiences, papers focusing on artifacts or benchmarks, and shorter work-in-progress papers.\footnote{Details about the call are available at \url{https://genaicreativity.org/icml2026/index.html\#CFP}} 
For the reviewing process, we assembled a program committee of $88$ reviewers, recruited through various channels: (a) sending invitations to researchers in our network, (b) releasing an open call for researchers with relevant experiences to act as reviewers, and (c) further adopting a reciprocal reviewing approach where one of the authors of each submission is requested to serve as a reviewer if need be.
We had $80$ paper submissions in total. The reviewing process was double-blind, and each paper received at least three reviews from our program committee.

The final workshop program included $49$ accepted papers.\footnote{Details about these papers can be found at \url{https://icml.cc/virtual/2026/workshop/54083}} 
The camera-ready PDFs of workshop papers are available as non-archival reports and are accessible on the workshop website along with the poster PDFs submitted by authors.\footnote{Link to camera-ready PDFs is at \url{https://genaicreativity.org/icml2026/index.html\#Papers}} Authors presented the papers in a poster session. As part of the workshop program, we also organized short lightning talks to provide authors with an opportunity to broaden the dissemination of their work. Figure~\ref{fig:word-cloud} illustrates a word cloud from titles of the papers. In Section~\ref{sec.research_directions}, we provide an overview of the key research topics covered in these papers at the workshop.

\begin{figure*}[h!]
    \vspace{2mm}
    \centering
    \includegraphics[trim={9cm 11cm 16cm 13cm},clip,width=1\textwidth]{./figs/word-cloud.pdf}
    \vspace{1mm}
    \caption{Word cloud (rendered as SEOUL) created from titles of accepted papers at the workshop.}
    \label{fig:word-cloud}
\end{figure*}

%

%
\subsection{Invited Talks and Panel Session}

We invited a set of speakers with diverse backgrounds, ranging from researchers with different expertise to industry professionals and practitioners, working in topics of generative AI, creativity, and human-AI co-creation.
In total, the workshop had $6$ invited talks; each of these talks being up to $30$ minutes long. 
The list of speakers, along with their talk titles and summaries of abstracts, is provided below:

\begin{itemize}[parsep=4pt,leftmargin=*,labelindent=6pt]
	\item  \href{https://people.cs.uchicago.edu/~ravenben/}{T1 Ben Y. Zhao}:  \texttalktitle{Challenges Facing Generative AI Users}. 
	The talk discussed several technical challenges that need to be addressed for generative content to be legitimized or adopted alongside human creative content.
	The talk also covered recent work highlighting AI slop in the music industry and the steps that must be taken before AI slop overwhelms legitimate musicians, including both human musicians and those using AI tools.
	References from the talk and related work by the speaker: \cite{DBLP:journals/corr/abs-2606-18052,DBLP:conf/uss/ShanCW0HZ23}.
 
	\item \href{https://elluba.com/about/}{T2 Luba Elliott}:  \texttalktitle{The Search for Originality in the Age of Mainstream AI}. 
    The talk discussed how the popularity of AI art has exploded over the past decade. The talk provided an overview of how artists have been using and thinking about AI, its creative potential, and its societal impact. The talk also presented examples of artworks by artists engaging with AI. References from the talk and related work by the speaker: \cite{cvprart2024}.

	\item \href{https://dxd-lab.github.io/}{T3  Hwajung Hong}:  \texttalktitle{Beyond the Average -- Orchestrating Human-AI Co-Creativity in Industrial Design}. 
	The talk discussed emerging issues because of growing reliance on generative AI tools in industrial design, including cognitive debt, homogenized outputs, and design fixation.
	The talk proposed two fundamental shifts to tackle these issues and improve human-AI teams for industrial design tasks: (a) transitioning from delegation to orchestration, where designers can command specialized agents grounded in human subjective judgment; (b) introducing intentional friction, such as casting an AI agent as a naive mentee to provoke critical thinking.
	References from the talk and related work by the speaker:  \cite{DBLP:journals/corr/abs-2506-08872,DBLP:conf/chi/Wadinambiarachchi24,DBLP:conf/chi/ShinPLO25,DBLP:conf/chi/ChaW25,DBLP:conf/chi/LimCNKH26,DBLP:journals/ijmms/LimCCNH26,DBLP:conf/iscid/DingCFLQC23}.
    
	\item \looseness-1 \href{https://johnr0.github.io/}{T4 John Joon Young Chung}:  \texttalktitle{Diversified and Explorative Language Models}.
    The talk discussed the limitations of large language models (LLMs) in supporting humans' creative, divergent, and exploratory tasks. In particular, LLM outputs tend to have limited diversity and humans tend to produce more homogenized results when assisted by LLMs. The talk presented two approaches to better align LLMs for these tasks: (a) a post-training approach that diversifies LLM outputs by learning more from rarer high-quality instances; (b) a user simulator that mimics humans' messy intentions during creative exploration and can be used to tune LLMs for explorative and creative tasks. 
    References from the talk and related work by the speaker: \cite{DBLP:conf/iclr/Padmakumar024,DBLP:conf/nips/JiangCLLFDTSC25,chung2025modifying,DBLP:journals/corr/abs-2602-03429}.
	\item \href{https://www.mindantix.com/team}{T5  Pronita Mehrotra}:  \texttalktitle{Creativity as Learning - Designing AI to Expand Human Thought}.
    The talk discussed how creativity and learning are two sides of the same coin, both involving some form of active reconstruction of internal mental models and generating novel possibilities. The talk presented how cognitive moves that support human creativity, such as associative and analogical thinking, can also improve the originality of AI-generated ideas. The talk highlighted that the greater potential of creativity research in AI lies in designing an AI system as a cognitive thought partner that helps humans strengthen higher-order thinking skills.
	References from the talk and related work by the speaker:  \cite{DBLP:journals/corr/abs-2405-06715,DBLP:journals/corr/abs-2512-18388,schwartz2004inventing,ugur2012effects}.

	\item \href{https://sites.google.com/site/lonnekenlp}{T6 Lonneke van der Plas}:  \texttalktitle{Modelling Language as a Vehicle for Creativity}. 
    The talk highlighted that creativity remains underexplored in NLP research, despite the ability of computational models of language to approximate some forms of human creativity. The talk motivated the need to investigate computational creativity and discussed specific challenges related to conducting research in creativity. The talk then presented several recent works, including modeling analogical reasoning and assessing/improving the creativity of outputs from large language models.
	References from the talk and related work by the speaker:  \cite{DBLP:conf/icccrea/LoiVP20,DBLP:journals/corr/abs-2006-11814,DBLP:journals/corr/abs-2605-06426,DBLP:conf/mwe/DharP19,DBLP:conf/emnlp/PetersenP23,DBLP:journals/tacl/PetersenSP26,DBLP:journals/corr/abs-2410-17218,DBLP:conf/icccrea/IsmayilzadaSP25,DBLP:conf/emnlp/IsmayilzadaLLPBPB25,DBLP:journals/corr/abs-2604-03480}.
	%

    %
\end{itemize} 

In addition to these invited talks, the speakers also participated in a panel session of $60$ minutes duration. The video recordings of these invited talks and the panel session are available online.\footnote{Link to video recordings can be found at \url{https://icml.cc/virtual/2026/workshop/54083}}

%

%

\section{Research Topics and Directions}
\label{sec.research_directions}
In this section, we provide an overview of popular research topics and highlight key research directions inspired by various workshop activities, including talks, panel discussions, and accepted papers. 
These research directions, along with the list of workshop papers closely related to each direction, are as follows: (1) Creativity Metrics -- see Section~\ref{direction-R1}~and~Table~\ref{fig:workshop-papers-R1}; (2) Benchmarks for Creative Tasks -- see Section~\ref{direction-R2}~and~Table~\ref{fig:workshop-papers-R2}; (3) Improving Creativity of Generative Models -- see Section~\ref{direction-R3}~and~Table~\ref{fig:workshop-papers-R3}; (4) Authenticity and Authorship of Creative Artifacts -- see Section~\ref{direction-R4}~and~Table~\ref{fig:workshop-papers-R4}; (5) Human-AI Co-Creation for Creative Tasks -- see Section~\ref{direction-R5}~and~Table~\ref{fig:workshop-papers-R5}.

\subsection{Creativity Metrics}\label{direction-R1}

\begin{table*}[th!]
    \centering
    \caption{Workshop papers related to the topic on creativity metrics.}
    \label{fig:workshop-papers-R1}   
    \scalebox{0.89}{
    \setlength\tabcolsep{1.5pt}
    \renewcommand{\arraystretch}{1.7}  
    \begin{tabular}{
        >{\centering\arraybackslash}p{0.1\textwidth} 
        >{\raggedright\arraybackslash}p{1.0\textwidth}
    }
        \toprule
        \textbf{Paper ID} & \multicolumn{1}{c}{\textbf{Paper Title}} \\
        \midrule
        \href{https://genaicreativity.org/icml2026/files/21/21_paper.pdf}{P$21$} & ``I've Seen How This Goes'': Characterizing the Diversity of LLM Generations and Human Writing via Progressive Conditional Surprise \\
        \href{https://genaicreativity.org/icml2026/files/26/26_paper.pdf}{P$26$} & Do Semantic Distance Tests Actually Predict Creativity in LLMs? \\
        \href{https://genaicreativity.org/icml2026/files/30/30_paper.pdf}{P$30$} & SEA: Understanding Sketch Abstraction Efficiency via Element-Level Commonsense for Human-AI Sketching \\
        \href{https://genaicreativity.org/icml2026/files/54/54_paper.pdf}{P$54$} & Rethinking Post-training Diversity Collapse: Is Diversity-preserving Post-training Enough? \\
        \href{https://genaicreativity.org/icml2026/files/55/55_paper.pdf}{P$55$} & Evaluating Design Video Generation: Metrics for Compositional Fidelity \\
        \bottomrule
    \end{tabular}
    } 
\end{table*}

\looseness-1There has been a long history of works seeking to define or measure creativity, though without a concise set of definitions that can be applied broadly across settings~\cite{DBLP:journals/corr/abs-2410-17218}. One useful framework is the model of \emph{four P's}~\cite{rhodes1961analysis}, which proposes evaluating creativity based on person, product, process, and press. 
Several classical tests have been proposed to measure the creativity of a person while also considering different stages of creative thinking processes: Alternative Uses Task (AUT) for divergent thinking \cite{guilford1956structure,guilford1967nature}, Remote Associates Test (RAT) for convergent thinking \cite{mednick1962associative}, and Torrance Tests of Creative Thinking (TTCT) as a collection of various creative thinking tasks~\cite{torrance1966torrance}.
In parallel, several different metrics have been proposed to capture different aspects of creativity for a product/artifact: novelty (i.e., originality or uniqueness), value (i.e., utility, effectiveness, or relevance), and surprise (i.e., non-obviousness)~\cite{runco2012standard,boden2004creative}.
There has also been work on developing measures to capture the environmental influence (i.e., ``press'' in the four P's); for instance, the Creativity Support Index (CSI) is a popular survey that is used to measure how an automated tool supports a user in terms of improving creativity~\cite{DBLP:journals/tochi/CherryL14}.
Recent works have investigated how generative AI models perform on these classical creativity tests in comparison to humans~\cite{hubert2024current,bellemare2026divergent}. There has also been a surge in interest in developing automated methods to compute these metrics to scale up evaluation and benchmarking of models~\cite{beaty2021automating, organisciak2023beyond,DBLP:journals/ia/FranceschelliM22,DBLP:conf/icalt/KarampiperisKK14}.
More recently, metrics have been proposed specifically for measuring the creativity of generative models~\cite{DBLP:conf/emnlp/QiuH25,DBLP:conf/iclr/LuSHM0HEJCD025}.
The invited talk [T6] at the workshop highlighted how various automated metrics compare with evaluations from human judges. Several workshop papers also investigated the computational aspects of creativity metrics, including the validity of semantic distance tests for creativity [P26], designing automated diversity measures [P21, P54], and domain-specific creativity metrics [P30, P55]  -- see Table~\ref{fig:workshop-papers-R1}.

Despite these recent works on designing or automating creativity metrics, a central challenge is the inherent subjectivity in evaluating various creativity dimensions -- there can be low agreement even across human annotations. Instead of dealing with individual creativity dimensions, an alternative evaluation method in the literature is the Consensual Assessment Technique~\cite{amabile1982social}, which proposes using a unified creativity rating provided by domain experts. In this regard, an interesting research direction is to develop automated methods to compute a unified creativity rating in a robust manner that aligns with domain experts while capturing different variations across individual experts.
Another issue is that standardized creativity tests or metrics designed for humans are not yet validated measures for generative AI models -- for instance, a model having a high or low score on an AUT or RAT test does not necessarily capture the creative capabilities of this model. An important research direction is to develop creativity metrics for generative AI models that are validated and calibrated with respect to human creativity.

\subsection{Benchmarks for Creative Tasks}\label{direction-R2}

\begin{table*}[th!]
    \centering
    \caption{Workshop papers related to the topic on benchmarks for creative tasks.}
    \label{fig:workshop-papers-R2}    
    \scalebox{0.89}{
    \setlength\tabcolsep{1.5pt}
    \renewcommand{\arraystretch}{1.7}  
    \begin{tabular}{
        >{\centering\arraybackslash}p{0.1\textwidth} 
        >{\raggedright\arraybackslash}p{1.0\textwidth}
    }
        \toprule
        \textbf{Paper ID} & \multicolumn{1}{c}{\textbf{Paper Title}} \\       
        \midrule
        \href{https://genaicreativity.org/icml2026/files/3/3_paper.pdf}{P$3$} & Beneath the Surface: Investigating LLMs' Capabilities for Communicating with Subtext \\
        \href{https://genaicreativity.org/icml2026/files/4/4_paper.pdf}{P$4$} & Multi-Genre Collection: Scaling Creative Writing Beyond Story-Centric Data \\
        \href{https://genaicreativity.org/icml2026/files/6/6_paper.pdf}{P$6$} & CresOWLve: Benchmarking Creative Problem-Solving Over Real-World Knowledge \\
        \href{https://genaicreativity.org/icml2026/files/16/16_paper.pdf}{P$16$} & A Typography Benchmark for Co-Creative Graphic-Design Agents \\
        \href{https://genaicreativity.org/icml2026/files/22/22_paper.pdf}{P$22$} & AutoFiction: Measuring AI ability to execute long-horizon writing tasks \\
        \href{https://genaicreativity.org/icml2026/files/32/32_paper.pdf}{P$32$} & Can Large Language Models Match Human Diversity in Educational Content Generation? \\
        \href{https://genaicreativity.org/icml2026/files/33/33_paper.pdf}{P$33$} & Object-Level Access to SVG Scenes: A Benchmark and Raster-Grounded Method for Text-Guided Vector Object Extraction \\
        \href{https://genaicreativity.org/icml2026/files/36/36_paper.pdf}{P$36$} & Visual Aesthetic Benchmark: Can Frontier Models Judge Beauty? \\
        \href{https://genaicreativity.org/icml2026/files/39/39_paper.pdf}{P$39$} & IDEAFix: Evaluation Framework for Creative Defixation Prompting in LLMs \\
        \href{https://genaicreativity.org/icml2026/files/44/44_paper.pdf}{P$44$} & SlideMatcher: Agentic Slide Editing under Item Mismatch \\
        \href{https://genaicreativity.org/icml2026/files/51/51_paper.pdf}{P$51$} & JudgeBreak: Is Taste Hidden in the Layers? Stress-Testing Reward Models and LLM Judges with Representation-Based Subjective Evaluation \\
        \href{https://genaicreativity.org/icml2026/files/57/57_paper.pdf}{P$57$} & GameDevBench: Evaluating Agentic Capabilities Through Game Development \\
        \href{https://genaicreativity.org/icml2026/files/60/60_paper.pdf}{P$60$} & NarrativeWorldBench: A Frontier-Saturated Benchmark and a Latent World Model for Long-Horizon Co-Creative Audio Drama \\
        \href{https://genaicreativity.org/icml2026/files/69/69_paper.pdf}{P$69$} & The Human Creativity Benchmark: Studying Convergence and Divergence in an Expert-Labeled Benchmark \\
        \href{https://genaicreativity.org/icml2026/files/77/77_paper.pdf}{P$77$} & Repairing Graphic Design Layouts After Content, Asset, and Canvas Updates: A Source-Conditioned Benchmark \\
        \bottomrule
    \end{tabular}
    }
\end{table*}

Existing works have reported a mismatch between how generative AI models are benchmarked and how they are used in everyday life~\cite{DBLP:conf/icml/Shen25, chatterji2025people}:  creative tasks are among the most prevalent scenarios when people interact with AI models, whereas benchmarking has primarily focused on generative models' accuracy or productivity.
To this end, recent works have proposed new benchmarks to evaluate model creativity. For instance, NoveltyBench~\cite{zhang2025noveltybench} measures models' ability to produce diverse, high-quality outputs to open-ended queries. Taking a step further, CreativityPrism~\cite{DBLP:journals/tmlr/HouZLBBLJBCCKL26} is a comprehensive benchmark comprising tasks such as divergent thinking, creative writing, and logical reasoning.
The invited talk [T6] also emphasized the importance of assessing the creativity of models -- also see paper [P6] from the speaker on benchmarking creative problem-solving.
Among the workshop papers, benchmarking was also one of the more popular topics and covered different domains, including writing \& language [P3, P4, P22, P32], visual/graphic design [P16, P33, P44, P77], ideation \& problem-solving [P6, P39], audio \& interactive media [P57, P60], and judgment [P36, P69] -- see Table~\ref{fig:workshop-papers-R2}. 
While these works seek to assess creativity in specific domains, the recent benchmarks tend to capture lower orders of creativity (``combinatorial'' or ``exploratory''~\cite{boden2004creative}) -- however, higher order of creativity (``transformational''~\cite{boden2004creative}) is more important in capturing model's capabilities to produce truly novel artifacts or capabilities for emerging application domains such as scientific discovery~\cite{gottweis2026accelerating}.
Another major shortcoming of these benchmarks is that they are designed to measure the creativity of a generative AI model when acting alone -- it remains unclear whether improving the capabilities of a model alone would necessarily transfer to improved human-AI co-creativity when a human user works along with such a model. To this end, an important research direction would be developing benchmarks specifically targeting human-AI co-creation settings.

\subsection{Improving Creativity of Generative Models}\label{direction-R3}

\begin{table*}[th!]
    \centering
    \caption{Workshop papers related to the topic on improving creativity of generative models.}
    \label{fig:workshop-papers-R3}  
    \scalebox{0.89}{
    \setlength\tabcolsep{1.5pt}
    \renewcommand{\arraystretch}{1.7}  
    \begin{tabular}{
        >{\centering\arraybackslash}p{0.1\textwidth} 
        >{\raggedright\arraybackslash}p{1.0\textwidth}
    }
        \toprule
        \textbf{Paper ID} & \multicolumn{1}{c}{\textbf{Paper Title}} \\
        \midrule
        \href{https://genaicreativity.org/icml2026/files/1/1_paper.pdf}{P$1$} & Caesar: Deep Agentic Web Exploration for Creative Answer Synthesis \\
        \href{https://genaicreativity.org/icml2026/files/7/7_paper.pdf}{P$7$} & Structured Creativity Methods for Multi-Agent LLMs: Brainwriting Outperforms Disney and Double Diamond on LLM-Judged Originality \\
        \href{https://genaicreativity.org/icml2026/files/9/9_paper.pdf}{P$9$} & The Missing Drive: Functional Analogs of Intrinsic Motivation in Large Language Models for Creative Tasks \\
        \href{https://genaicreativity.org/icml2026/files/11/11_paper.pdf}{P$11$} & Adrasteia: A Human-in-the-Loop Agentic Workflow for Theory-Oriented Research in Economics \\
        \href{https://genaicreativity.org/icml2026/files/12/12_paper.pdf}{P$12$} & CreativityNeuro: Steering Language Model Weights to Improve Divergent Thinking and Reduce Mode Collapse \\
        \href{https://genaicreativity.org/icml2026/files/13/13_paper.pdf}{P$13$} & Seeking the Unfamiliar but Memorable: Conceptual Creativity as Meta-Learning \\
        \href{https://genaicreativity.org/icml2026/files/17/17_paper.pdf}{P$17$} & Optimizing Diversity and Quality through Base-Aligned Model Collaboration \\
        \href{https://genaicreativity.org/icml2026/files/40/40_paper.pdf}{P$40$} & VendiEvolve: Towards Diversity-Aware LLM-Guided Evolutionary Search \\
        \href{https://genaicreativity.org/icml2026/files/62/62_paper.pdf}{P$62$} & We Built a Salon, We Called It Alignment: Aesthetic Reward and the Latent Avant-Garde in Generative Models \\
        \href{https://genaicreativity.org/icml2026/files/63/63_paper.pdf}{P$63$} & Süssmayr's Heirs: A Composition-Invariant View of Generative Stylistic Completion \\
        \href{https://genaicreativity.org/icml2026/files/67/67_paper.pdf}{P$67$} & On the Transfer of Output Diversity via Synthetic Data in Language Models \\
        \href{https://genaicreativity.org/icml2026/files/71/71_paper.pdf}{P$71$} & ArtMine: Discovering and Formalizing Artistic Processes \\
        \href{https://genaicreativity.org/icml2026/files/72/72_paper.pdf}{P$72$} & Forty Shades of Blue: Quality-Diversity Alignment via Mode-Conditioned Reinforcement Learning \\
        \href{https://genaicreativity.org/icml2026/files/78/78_paper.pdf}{P$78$} & Creative Collision: Directorial Persona Steering and Competition in Large Language Models \\
        \bottomrule
    \end{tabular}
    }  
\end{table*}

Existing works have shown that generative AI models exhibit an ``Artificial Hivemind'' effect, i.e., models tend to produce homogenized outputs both within a model and across models \cite{DBLP:conf/nips/JiangCLLFDTSC25}. This homogenization is further reinforced by post-training methods, which typically improve task accuracy at the cost of output diversity
\cite{DBLP:conf/iclr/KirkMNLHGR24}. Recent works have investigated different approaches to enhance the diversity and creativity of model outputs, both at inference time~\cite{zhang2025noveltybench,2026.EDM.poster-demo-papers.424} and at training time using signals from multiple creativity dimensions \cite{DBLP:conf/emnlp/IsmayilzadaLLPBPB25,chung2025modifying}. Several invited talks at the workshop focused on this topic of improving the diversity or creativity of generative models (see talks [T4, T5, T6]). Moreover, several workshop papers also proposed new methods that intervene at different points of the training or inference stages, including steering model activations [P12, P78], decoding by combining models [P17], better training data [P67], and new training objectives [P72] -- see Table~\ref{fig:workshop-papers-R3}. 

These works show the promise of improving the diversity and creativity of generative AI models, though the key challenge is how to achieve this objective while maintaining the quality of models' outputs.
There have been recent efforts in tackling this question from the perspective of set-level reinforcement learning, where a model is trained to optimize a diversity-aware set function defined over rollouts~\cite{hamid2026polychromic}. An interesting research direction would be to explore the question of improving diversity/creativity of a model from the lens of multi-agent policy architectures, which can provide formalism for attributing rewards to different rollouts when optimizing a set function during training.
Another exciting and important direction for future work is designing new architectures for creativity-centric generative models by leveraging insights from the literature on human creativity~\cite{finke1992creative,guilford1967nature} -- for instance, a generative model with dedicated divergent-convergent thinking phases before outputting an answer.
Another issue is that these works typically aim to improve the model's diversity or creativity when acting alone without considering human users in co-creation settings. An important research direction for future work is designing new architectures and methods that better support human-AI co-creation and boost human-AI co-creativity.

\subsection{Authenticity and Authorship of Creative Artifacts}\label{direction-R4}

\begin{table*}[th!]
    \centering
    \caption{Workshop papers related to the topic on authenticity and authorship of creative artifacts.}
    \label{fig:workshop-papers-R4}  
    \scalebox{0.89}{
    \setlength\tabcolsep{1.5pt}
    \renewcommand{\arraystretch}{1.7}  
    \begin{tabular}{
        >{\centering\arraybackslash}p{0.1\textwidth} 
        >{\raggedright\arraybackslash}p{1.0\textwidth}
    }
        \toprule
        \textbf{Paper ID} & \multicolumn{1}{c}{\textbf{Paper Title}} \\
        \midrule
        \href{https://genaicreativity.org/icml2026/files/46/46_paper.pdf}{P$46$} & When Machines Decide If a Human Wrote It: Creativity in the Age of AI Detectors \\
        \href{https://genaicreativity.org/icml2026/files/70/70_paper.pdf}{P$70$} & ``I Didn't Make the Micro Decisions'': Measuring, Inducing, and Exposing Goal-Level AI Contributions in Collaboration \\
        \href{https://genaicreativity.org/icml2026/files/80/80_paper.pdf}{P$80$} & Scalable and Interpretable Authorship Attribution for AI Generated Code \\
        \bottomrule
    \end{tabular}
    }  
\end{table*}

\looseness-1The integration of generative AI agents into creative workflows has raised serious issues of content copyright, authorship, and longer-term consequences for creative professions~\cite{samuelson2023generative,DBLP:conf/aies/JiangBCKGWHFG23,DBLP:conf/aies/LovatoZSDK24}.
The invited talk [T1] at the workshop highlighted these issues and also discussed technical challenges that need to be addressed for generative content to be legitimized or adopted alongside human creative content. The talk also discussed two works: the Glaze tool to protect artists from style mimicry by generative models~\cite{DBLP:conf/uss/ShanCW0HZ23} and recent findings on AI slop in the music industry~\cite{DBLP:journals/corr/abs-2606-18052}.
Several recent works have also examined AI-content detection~\cite{DBLP:journals/tmlr/SadasivanKB0F25,DBLP:journals/patterns/LiangYMWZ23}, contribution attribution~\cite{DBLP:conf/chi/HeHW25}, watermarking~\cite{DBLP:conf/icml/KirchenbauerGWK23,DBLP:journals/nature/DathathriSGHMWBKSMHVMBB24}, and creator safeguarding~\cite{DBLP:conf/uss/ShanCW0HZ23,DBLP:conf/imc/0001LSVZS25}. 
Several workshop papers explored methods to address these issues, including the consequences of using AI detectors on human creative behavior [P46], how AI can influence users' goals and micro-decisions [P70], and authorship attribution for AI-generated code [P80].

\looseness-1A central challenge is that as generative AI agents become more capable across content modalities, methods designed for authenticity and authorship must be continuously evaluated and improved.
Another challenge arises from the ease and scale at which AI content can be generated and uploaded online (e.g., see the scale of AI-generated music reported in~\cite{DBLP:journals/corr/abs-2606-18052}). Tackling this challenge would require developing detection and attribution methods that operate robustly at internet scale. 
Another important research direction is to develop new process-level provenance mechanisms that capture who contributed what and to what extent over time. This fine-grained provenance is important both within a co-creation session and across sessions as creative content gets generated or remixed over time.

\subsection{Human-AI Co-Creation for Creative Tasks}
\label{direction-R5}

\begin{table*}[th!]
    \centering
    \caption{Workshop papers related to the topic on human-AI co-creation for creative tasks.}
    \label{fig:workshop-papers-R5}  
    \scalebox{0.89}{
    \setlength\tabcolsep{1.5pt}
    \renewcommand{\arraystretch}{1.7}  
    \begin{tabular}{
        >{\centering\arraybackslash}p{0.1\textwidth} 
        >{\raggedright\arraybackslash}p{1.0\textwidth}
    }
        \toprule
        \textbf{Paper ID} & \multicolumn{1}{c}{\textbf{Paper Title}} \\
        \midrule
        \href{https://genaicreativity.org/icml2026/files/19/19_paper.pdf}{P$19$} & Incentives shape how humans co-create with generative AI \\
        \href{https://genaicreativity.org/icml2026/files/24/24_paper.pdf}{P$24$} & Feedback-to-Rubrics: Can We Extract Expert Criteria from Inline Comments? \\
        \href{https://genaicreativity.org/icml2026/files/29/29_paper.pdf}{P$29$} & Bonsai: Cultivating Author Intent in LLM-Based Interactive Digital Narratives \\
        \href{https://genaicreativity.org/icml2026/files/31/31_paper.pdf}{P$31$} & Trajectory as Creative Intent: Rethinking Drag-Based Interaction in Generative Visual Editing \\
        \href{https://genaicreativity.org/icml2026/files/38/38_paper.pdf}{P$38$} & Bridging Creative Intent and Visual Quality: Creator-Driven Recurrent Video Generation with Agentic Feedback Loops \\
        \href{https://genaicreativity.org/icml2026/files/43/43_paper.pdf}{P$43$} & SVG-BERT: Code-Native Representations for Structure-Aware Vector Graphics Workflows \\
        \href{https://genaicreativity.org/icml2026/files/49/49_paper.pdf}{P$49$} & Creativity from Friction: Human-AI Interaction for Exploratory Structural Design \\
        \href{https://genaicreativity.org/icml2026/files/53/53_paper.pdf}{P$53$} & User Engagement With Intermediate Representations in LLM-Assisted Audio Programming \\
        \href{https://genaicreativity.org/icml2026/files/73/73_paper.pdf}{P$73$} & Searching for Synergy in Shared Workspace Human-AI Collaboration \\
        \href{https://genaicreativity.org/icml2026/files/81/81_paper.pdf}{P$81$} & Toward User-Orchestrated Human-Multi-Agent Teams for Co-Creation \\
        \bottomrule
    \end{tabular}
    }  
\end{table*}

\looseness-1
Existing works have shown that AI-generated suggestions or solutions can fixate or narrow users' thinking, leading to issues such as design fixation~\cite{DBLP:conf/chi/Wadinambiarachchi24} and over-reliance~\cite{DBLP:journals/corr/abs-2506-08872,DBLP:journals/bjet/FanTLSTZSLG25,DBLP:conf/chi/LeeSTDRBW25}. 
To address the design fixation issue, recent works have proposed new exploration mechanisms for users, such as using structured design-space exploration~\cite{DBLP:conf/chi/SuhCMLX24} and 
surfacing ideas that users might not reach on their own~\cite{DBLP:journals/corr/abs-2512-18388}. To reduce over-reliance on AI, there are several interaction paradigms proposed in the literature that could be useful, such as introducing cognitive forcing interventions~\cite{DBLP:journals/pacmhci/BucincaMG21}, the use of AI explanations to support verification~\cite{DBLP:journals/pacmhci/VasconcelosJGGBK23}, and having a user reflect on their work before seeking AI assistance~\cite{DBLP:journals/corr/abs-2512-04630}.
In the workshop, the invited talk [T3] discussed the challenges of growing reliance on generative AI and presented several ways to address them, e.g., by focusing human-AI interaction on orchestration and introducing intentional friction (e.g., casting an AI agent as a naive mentee to provoke critical thinking). The invited talk [T5] also highlighted that the greater potential of creativity-centered AI systems is to act as a cognitive thought partner that helps humans strengthen higher-order thinking skills.
Several workshop papers also investigated approaches to improve human-AI co-creation for creative tasks, including designing incentives to shape creative diversity [P19], mechanisms for preserving creative intent [P29, P31, P38], introducing friction for creativity [P49], and introducing human orchestration for creative interaction [P81].

One important research direction is to develop rigorous engagement measures that capture design fixation and over-reliance throughout the creative process and also generalize across domains. These measures can then support interaction mechanisms with active interventions to address these issues during a human-AI co-creation session. 
Another issue is that most of the above-mentioned works focus on boosting human-AI co-creativity at the level of individual users; however, existing studies have highlighted the risk that enhancing individual-level creativity can still harm collective diversity across a population of users~\cite {doi:10.1126/sciadv.adn5290}. To this end, an important research challenge is how to design generative AI systems that boost both individual-level and population-level creativity.
Another exciting research direction is how to design AI systems that 
go beyond focusing on final artifacts and also optimize for dynamic effects such as user agency or transparency within a co-creation session, as well as user learning that transfers across sessions.

%


\vspace{-1mm}
\section{Conclusions}
\label{sec.conclusions}
\vspace{-1mm}
Human-AI co-creativity is an emerging, time-critical area for researchers, industry professionals, and practitioners: ongoing advances in generative AI will continue to provide unprecedented opportunities to support people in open-ended domains, and the unique challenges of integrating generative AI agents into creative workflows will require new safeguards along with technical innovations.
The talks, papers, and discussions at the ICML'26 workshop highlighted the community's excitement around the main topics covered in this article, and the diverse array of perspectives demonstrated the importance of drawing on ideas from multiple disciplines.
This need for diverse perspectives and expertise highlights the need to continue fostering collaboration among different communities and stakeholders in the area of human-AI co-creativity.

%

\bibliography{main}

\begin{thebibliography}{10}

\bibitem{DBLP:journals/corr/abs-2606-18052}
Stanley Wu, Josephine Passananti, Viresh Mittal, Wenxin Ding, Hai{-}Tao Zheng, and Ben~Y. Zhao.
\newblock {An Empirical Analysis of {AI} Slop in Music Streaming}.
\newblock {\em CoRR}, abs/2606.18052, 2026.
\newblock \href {https://doi.org/10.48550/ARXIV.2606.18052} {\path{doi:10.48550/ARXIV.2606.18052}}.

\bibitem{DBLP:conf/uss/ShanCW0HZ23}
Shawn Shan, Jenna Cryan, Emily Wenger, Haitao Zheng, Rana Hanocka, and Ben~Y. Zhao.
\newblock {Glaze: Protecting Artists from Style Mimicry by Text-to-Image Models}.
\newblock In {\em {USENIX} Security Symposium, {USENIX} Security}, pages 2187--2204, 2023.
\newblock URL: \url{https://www.usenix.org/conference/usenixsecurity23/presentation/shan}.

\bibitem{cvprart2024}
CVPR Art.
\newblock {CVPR 2024 Art Gallery}.
\newblock \url{https://thecvf-art.com/archive.php?year=2024}, 2024.

\bibitem{DBLP:journals/corr/abs-2506-08872}
Nataliya Kosmyna, Eugene Hauptmann, Ye~Tong Yuan, Jessica Situ, Xian{-}Hao Liao, Ashly~Vivian Beresnitzky, Iris Braunstein, and Pattie Maes.
\newblock {Your Brain on ChatGPT: Accumulation of Cognitive Debt when Using an {AI} Assistant for Essay Writing Task}.
\newblock {\em CoRR}, abs/2506.08872, 2025.
\newblock \href {https://doi.org/10.48550/ARXIV.2506.08872} {\path{doi:10.48550/ARXIV.2506.08872}}.

\bibitem{DBLP:conf/chi/Wadinambiarachchi24}
Samangi Wadinambiarachchi, Ryan~M. Kelly, Saumya Pareek, Qiushi Zhou, and Eduardo Velloso.
\newblock {The Effects of Generative {AI} on Design Fixation and Divergent Thinking}.
\newblock In {\em Proceedings of the {CHI} Conference on Human Factors in Computing Systems ({CHI})}, pages 380:1--380:18, 2024.
\newblock \href {https://doi.org/10.1145/3613904.3642919} {\path{doi:10.1145/3613904.3642919}}.

\bibitem{DBLP:conf/chi/ShinPLO25}
Joongi Shin, Anna Polyanskaya, Andr{\'{e}}s Lucero, and Antti Oulasvirta.
\newblock {No Evidence for LLMs Being Useful in Problem Reframing}.
\newblock In {\em Proceedings of the {CHI} Conference on Human Factors in Computing Systems ({CHI})}, pages 243:1--243:25, 2025.
\newblock \href {https://doi.org/10.1145/3706598.3713273} {\path{doi:10.1145/3706598.3713273}}.

\bibitem{DBLP:conf/chi/ChaW25}
Inha Cha and Richmond~Y. Wong.
\newblock {Understanding Socio-technical Factors Configuring {AI} Non-Use in {UX} Work Practices}.
\newblock In {\em Proceedings of the {CHI} Conference on Human Factors in Computing Systems ({CHI})}, pages 1110:1--1110:17, 2025.
\newblock \href {https://doi.org/10.1145/3706598.3713140} {\path{doi:10.1145/3706598.3713140}}.

\bibitem{DBLP:conf/chi/LimCNKH26}
Hyunseung Lim, Dasom Choi, Sooyohn Nam, Bogoan Kim, and Hwajung Hong.
\newblock {Understanding Human-Multi-Agent Team Formation for Creative Work}.
\newblock In {\em Proceedings of the {CHI} Conference on Human Factors in Computing Systems ({CHI})}, pages 70:1--70:21, 2026.
\newblock \href {https://doi.org/10.1145/3772318.3791166} {\path{doi:10.1145/3772318.3791166}}.

\bibitem{DBLP:journals/ijmms/LimCCNH26}
Hyunseung Lim, Dasom Choi, DaEun Choi, Sooyohn Nam, and Hwajung Hong.
\newblock {Feed-O-Meter: Investigating AI-generated Mentee Personas as Interactive Agents for Scaffolding Design Feedback Practice}.
\newblock {\em International Journal of Human-Computer Studies}, 208:103687, 2026.
\newblock \href {https://doi.org/10.1016/J.IJHCS.2025.103687} {\path{doi:10.1016/J.IJHCS.2025.103687}}.

\bibitem{DBLP:conf/iscid/DingCFLQC23}
Shiying Ding, Xinyi Chen, Yan Fang, Wenrui Liu, Yiwu Qiu, and Chunlei Chai.
\newblock {DesignGPT: Multi-Agent Collaboration in Design}.
\newblock In {\em International Symposium on Computational Intelligence and Design ({ISCID})}, pages 204--208, 2023.
\newblock \href {https://doi.org/10.1109/ISCID59865.2023.00056} {\path{doi:10.1109/ISCID59865.2023.00056}}.

\bibitem{DBLP:conf/iclr/Padmakumar024}
Vishakh Padmakumar and He~He.
\newblock {Does Writing with Language Models Reduce Content Diversity?}
\newblock In {\em International Conference on Learning Representations ({ICLR})}, 2024.
\newblock URL: \url{https://openreview.net/forum?id=Feiz5HtCD0}.

\bibitem{DBLP:conf/nips/JiangCLLFDTSC25}
Liwei Jiang, Yuanjun Chai, Margaret Li, Mickel Liu, Raymond Fok, Nouha Dziri, Yulia Tsvetkov, Maarten Sap, and Yejin Choi.
\newblock {Artificial Hivemind: The Open-Ended Homogeneity of Language Models (and Beyond)}.
\newblock In {\em Annual Conference on Neural Information Processing Systems ({NeurIPS})}, 2025.
\newblock URL: \url{http://papers.nips.cc/paper\_files/paper/2025/hash/754d5a526a5ee5a47220664a0eb92751-Abstract-Datasets\_and\_Benchmarks\_Track.html}.

\bibitem{chung2025modifying}
John Joon~Young Chung, Vishakh Padmakumar, Melissa Roemmele, Yuqian Sun, and Max Kreminski.
\newblock {Modifying Large Language Model Post-Training for Diverse Creative Writing}.
\newblock In {\em Conference on Language Modeling (COLM)}, 2025.
\newblock URL: \url{https://openreview.net/forum?id=1Pmuw08LoM}.

\bibitem{DBLP:journals/corr/abs-2602-03429}
Tae~Soo Kim, Yoonjoo Lee, Jaesang Yu, John Joon~Young Chung, and Juho Kim.
\newblock {DiscoverLLM: From Executing Intents to Discovering Them}.
\newblock {\em CoRR}, abs/2602.03429, 2026.
\newblock \href {https://doi.org/10.48550/ARXIV.2602.03429} {\path{doi:10.48550/ARXIV.2602.03429}}.

\bibitem{DBLP:journals/corr/abs-2405-06715}
Pronita Mehrotra, Aishni Parab, and Sumit Gulwani.
\newblock {Enhancing Creativity in Large Language Models through Associative Thinking Strategies}.
\newblock {\em CoRR}, abs/2405.06715, 2024.
\newblock \href {https://doi.org/10.48550/ARXIV.2405.06715} {\path{doi:10.48550/ARXIV.2405.06715}}.

\bibitem{DBLP:journals/corr/abs-2512-18388}
Chao Wen, Tung Phung, Pronita Mehrotra, Sumit Gulwani, Roger~E. Beaty, Tomohiro Nagashima, and Adish Singla.
\newblock {Exploration vs. Fixation: Scaffolding Divergent and Convergent Thinking for Human-AI Co-Creation with Generative Models}.
\newblock {\em CoRR}, abs/2512.18388, 2025.
\newblock \href {https://doi.org/10.48550/ARXIV.2512.18388} {\path{doi:10.48550/ARXIV.2512.18388}}.

\bibitem{schwartz2004inventing}
Daniel~L Schwartz and Taylor Martin.
\newblock {Inventing to Prepare for Future Learning: The Hidden Efficiency of Encouraging Original Student Production in Statistics Instruction}.
\newblock {\em Cognition and Instruction}, 22(2):129--184, 2004.
\newblock \href {https://doi.org/10.1207/s1532690xci2202_1} {\path{doi:10.1207/s1532690xci2202_1}}.

\bibitem{ugur2012effects}
Gokhan Ugur, Refik Dilber, Yasemin Senpolat, and Bahattin Duzgun.
\newblock {The Effects of Analogy on Students' Understanding of Direct Current Circuits and Attitudes Towards Physics Lessons}.
\newblock {\em European Journal of Educational Research}, 1(3):211--223, 2012.
\newblock \href {https://doi.org/10.12973/eu-jer.1.3.211} {\path{doi:10.12973/eu-jer.1.3.211}}.

\bibitem{DBLP:conf/icccrea/LoiVP20}
Michele Loi, Eleonora Vigan{\`{o}}, and Lonneke van~der Plas.
\newblock {The Societal and Ethical Relevance of Computational Creativity}.
\newblock In {\em Proceedings of the International Conference on Computational Creativity ({ICCC})}, pages 398--401, 2020.
\newblock URL: \url{https://computationalcreativity.net/iccc20/papers/160-iccc20.pdf}.

\bibitem{DBLP:journals/corr/abs-2006-11814}
Michele Loi and Lonneke van~der Plas.
\newblock {A Blindspot of {AI} Ethics: Anti-fragility in Statistical Prediction}.
\newblock {\em CoRR}, abs/2006.11814, 2020.
\newblock URL: \url{https://arxiv.org/abs/2006.11814}.

\bibitem{DBLP:journals/corr/abs-2605-06426}
Diego Rossini and Lonneke van~der Plas.
\newblock {From 124 Million Tokens to 1,021 Neologisms: {A} Large-Scale Pipeline for Automatic Neologism Detection}.
\newblock {\em CoRR}, abs/2605.06426, 2026.
\newblock \href {https://doi.org/10.48550/ARXIV.2605.06426} {\path{doi:10.48550/ARXIV.2605.06426}}.

\bibitem{DBLP:conf/mwe/DharP19}
Prajit Dhar and Lonneke van~der Plas.
\newblock {Learning to Predict Novel Noun-Noun Compounds}.
\newblock In {\em Proceedings of the Joint Workshop on Multiword Expressions and WordNet (MWE-WN@ACL)}, pages 30--39, 2019.
\newblock \href {https://doi.org/10.18653/V1/W19-5105} {\path{doi:10.18653/V1/W19-5105}}.

\bibitem{DBLP:conf/emnlp/PetersenP23}
Molly~R. Petersen and Lonneke van~der Plas.
\newblock {Can Language Models Learn Analogical Reasoning? Investigating Training Objectives and Comparisons to Human Performance}.
\newblock In {\em Proceedings of the Conference on Empirical Methods in Natural Language Processing, ({EMNLP})}, pages 16414--16425, 2023.
\newblock \href {https://doi.org/10.18653/V1/2023.EMNLP-MAIN.1022} {\path{doi:10.18653/V1/2023.EMNLP-MAIN.1022}}.

\bibitem{DBLP:journals/tacl/PetersenSP26}
Molly~R. Petersen, Claire~E. Stevenson, and Lonneke van~der Plas.
\newblock {Modelling Analogies and Analogical Reasoning: Connecting Cognitive Science Theory and {NLP} Research}.
\newblock {\em Transactions of the Association for Computational Linguistics}, 14:711--732, 2026.
\newblock \href {https://doi.org/10.1162/TACL.A.632} {\path{doi:10.1162/TACL.A.632}}.

\bibitem{DBLP:journals/corr/abs-2410-17218}
Mete Ismayilzada, Debjit Paul, Antoine Bosselut, and Lonneke van~der Plas.
\newblock {Creativity in {AI:} Progresses and Challenges}.
\newblock {\em CoRR}, abs/2410.17218, 2024.
\newblock \href {https://doi.org/10.48550/ARXIV.2410.17218} {\path{doi:10.48550/ARXIV.2410.17218}}.

\bibitem{DBLP:conf/icccrea/IsmayilzadaSP25}
Mete Ismayilzada, Claire~E. Stevenson, and Lonneke van~der Plas.
\newblock {Evaluating Creative Short Story Generation in Humans and Large Language Models}.
\newblock In {\em Proceedings of the International Conference on Computational Creativity (ICCC)}, pages 145--155, 2025.
\newblock URL: \url{https://computationalcreativity.net/iccc25/papers/iccc25-ismayilzada2025evaluating.pdf}.

\bibitem{DBLP:conf/emnlp/IsmayilzadaLLPBPB25}
Mete Ismayilzada, Antonio~Laverghetta Jr., Simone Luchini, Reet Patel, Antoine Bosselut, Lonneke van~der Plas, and Roger~E. Beaty.
\newblock {Creative Preference Optimization}.
\newblock In {\em Findings of the Association for Computational Linguistics: {EMNLP} ({EMNLP Findings})}, pages 9580--9609, 2025.
\newblock \href {https://doi.org/10.18653/V1/2025.FINDINGS-EMNLP.509} {\path{doi:10.18653/V1/2025.FINDINGS-EMNLP.509}}.

\bibitem{DBLP:journals/corr/abs-2604-03480}
Mete Ismayilzada, Simone Luchini, Abdulkadir Gokce, Badr AlKhamissi, Antoine Bosselut, Antonio~Laverghetta Jr., Lonneke van~der Plas, and Roger~E. Beaty.
\newblock {Large Language Models Align with the Human Brain during Creative Thinking}.
\newblock {\em CoRR}, abs/2604.03480, 2026.
\newblock \href {https://doi.org/10.48550/ARXIV.2604.03480} {\path{doi:10.48550/ARXIV.2604.03480}}.

\bibitem{rhodes1961analysis}
Mel Rhodes.
\newblock {An Analysis of Creativity}.
\newblock {\em The Phi Delta Kappan}, 42(7):305--310, 1961.
\newblock URL: \url{https://www.jstor.org/stable/20342603}.

\bibitem{guilford1956structure}
Joy~Paul Guilford.
\newblock {The Structure of Intellect.}
\newblock {\em Psychological bulletin}, 53(4):267, 1956.
\newblock \href {https://doi.org/10.1037/h0040755} {\path{doi:10.1037/h0040755}}.

\bibitem{guilford1967nature}
Joy~Paul Guilford.
\newblock {The Nature of Human Intelligence}.
\newblock 1967.
\newblock URL: \url{https://psycnet.apa.org/record/1967-35015-000}.

\bibitem{mednick1962associative}
Sarnoff Mednick.
\newblock {The Associative Basis of The Creative Process.}
\newblock {\em Psychological Review}, 69(3):220, 1962.
\newblock \href {https://doi.org/10.1037/h0048850} {\path{doi:10.1037/h0048850}}.

\bibitem{torrance1966torrance}
E~Paul Torrance.
\newblock {Torrance Tests of Creative Thinking}.
\newblock {\em Educational and Psychological Measurement}, 1966.
\newblock \href {https://doi.org/10.1037/t05532-000} {\path{doi:10.1037/t05532-000}}.

\bibitem{runco2012standard}
Mark~A Runco and Garrett~J Jaeger.
\newblock {The Standard Definition of Creativity}.
\newblock {\em Creativity Research Journal}, 24(1):92--96, 2012.
\newblock \href {https://doi.org/10.1080/10400419.2012.650092} {\path{doi:10.1080/10400419.2012.650092}}.

\bibitem{boden2004creative}
Margaret~A Boden.
\newblock {\em {The Creative Mind: Myths and Mechanisms}}.
\newblock Routledge, 2004.
\newblock \href {https://doi.org/10.4324/9780203508527} {\path{doi:10.4324/9780203508527}}.

\bibitem{DBLP:journals/tochi/CherryL14}
Erin Cherry and Celine Latulipe.
\newblock {Quantifying the Creativity Support of Digital Tools through the Creativity Support Index}.
\newblock {\em {ACM Transactions on Computer-Human Interaction}}, 21(4):21:1--21:25, 2014.
\newblock \href {https://doi.org/10.1145/2617588} {\path{doi:10.1145/2617588}}.

\bibitem{hubert2024current}
Kent~F Hubert, Kim~N Awa, and Darya~L Zabelina.
\newblock {The Current State of Artificial Intelligence Generative Language Models is More Creative than Humans on Divergent Thinking Tasks}.
\newblock {\em Scientific Reports}, 14(1):3440, 2024.
\newblock \href {https://doi.org/10.1038/s41598-024-53303-w} {\path{doi:10.1038/s41598-024-53303-w}}.

\bibitem{bellemare2026divergent}
Antoine Bellemare-Pepin, Fran{\c{c}}ois Lespinasse, Philipp Th{\"o}lke, Yann Harel, Kory Mathewson, Jay~A Olson, Yoshua Bengio, and Karim Jerbi.
\newblock {Divergent Creativity in Humans and Large Language Models}.
\newblock {\em Scientific Reports}, 16(1):1279, 2026.
\newblock \href {https://doi.org/10.1038/s41598-025-25157-3} {\path{doi:10.1038/s41598-025-25157-3}}.

\bibitem{beaty2021automating}
Roger~E Beaty and Dan~R Johnson.
\newblock {Automating Creativity Assessment with SemDis: An Open Platform for Computing Semantic Distance}.
\newblock {\em Behavior Research Methods}, 53(2):757--780, 2021.
\newblock \href {https://doi.org/10.3758/s13428-020-01453-w} {\path{doi:10.3758/s13428-020-01453-w}}.

\bibitem{organisciak2023beyond}
Peter Organisciak, Selcuk Acar, Denis Dumas, and Kelly Berthiaume.
\newblock {Beyond Semantic Distance: Automated Scoring of Divergent Thinking Greatly Improves with Large Language Models}.
\newblock {\em Thinking Skills and Creativity}, 49:101356, 2023.
\newblock \href {https://doi.org/10.1016/j.tsc.2023.101356} {\path{doi:10.1016/j.tsc.2023.101356}}.

\bibitem{DBLP:journals/ia/FranceschelliM22}
Giorgio Franceschelli and Mirco Musolesi.
\newblock {DeepCreativity: Measuring Creativity with Deep Learning Techniques}.
\newblock {\em Intelligenza Artificiale}, 16(2):151--163, 2022.
\newblock \href {https://doi.org/10.3233/IA-220136} {\path{doi:10.3233/IA-220136}}.

\bibitem{DBLP:conf/icalt/KarampiperisKK14}
Pythagoras Karampiperis, Antonis Koukourikos, and Evangelia Koliopoulou.
\newblock {Towards Machines for Measuring Creativity: The Use of Computational Tools in Storytelling Activities}.
\newblock In {\em International Conference on Advanced Learning Technologies ({ICALT})}, pages 508--512, 2014.
\newblock \href {https://doi.org/10.1109/ICALT.2014.150} {\path{doi:10.1109/ICALT.2014.150}}.

\bibitem{DBLP:conf/emnlp/QiuH25}
Ziliang Qiu and Renfen Hu.
\newblock {Deep Associations, High Creativity: {A} Simple yet Effective Metric for Evaluating Large Language Models}.
\newblock In {\em Proceedings of the Conference on Empirical Methods in Natural Language Processing, ({EMNLP})}, pages 10859--10872, 2025.
\newblock \href {https://doi.org/10.18653/V1/2025.EMNLP-MAIN.550} {\path{doi:10.18653/V1/2025.EMNLP-MAIN.550}}.

\bibitem{DBLP:conf/iclr/LuSHM0HEJCD025}
Ximing Lu, Melanie Sclar, Skyler Hallinan, Niloofar Mireshghallah, Jiacheng Liu, Seungju Han, Allyson Ettinger, Liwei Jiang, Khyathi~Raghavi Chandu, Nouha Dziri, and Yejin Choi.
\newblock {{AI} as Humanity's Salieri: Quantifying Linguistic Creativity of Language Models via Systematic Attribution of Machine Text against Web Text}.
\newblock In {\em International Conference on Learning Representations ({ICLR})}, 2025.
\newblock URL: \url{https://openreview.net/forum?id=ilOEOIqolQ}.

\bibitem{amabile1982social}
Teresa~M Amabile.
\newblock {Social Psychology of Creativity: A Consensual Assessment Technique.}
\newblock {\em Journal of Personality and Social Psychology}, 43(5):997, 1982.
\newblock \href {https://doi.org/10.1037/0022-3514.43.5.997} {\path{doi:10.1037/0022-3514.43.5.997}}.

\bibitem{DBLP:conf/icml/Shen25}
Judy~Hanwen Shen.
\newblock {Position: Societal Impacts Research Requires Benchmarks for Creative Composition Tasks}.
\newblock In {\em International Conference on Machine Learning ({ICML})}, 2025.
\newblock URL: \url{https://proceedings.mlr.press/v267/shen25r.html}.

\bibitem{chatterji2025people}
Aaron Chatterji, Thomas Cunningham, David~J Deming, Zoe Hitzig, Christopher Ong, Carl~Yan Shan, and Kevin Wadman.
\newblock {How People Use ChatGPT}.
\newblock Technical report, {National Bureau of Economic Research}, 2025.
\newblock \href {https://doi.org/10.3386/w34255} {\path{doi:10.3386/w34255}}.

\bibitem{zhang2025noveltybench}
Yiming Zhang, Harshita Diddee, Susan Holm, Hanchen Liu, Xinyue Liu, Vinay Samuel, Barry Wang, and Daphne Ippolito.
\newblock {NoveltyBench: Evaluating Creativity and Diversity in Language Models}.
\newblock In {\em Conference on Language Modeling (COLM)}, 2025.
\newblock URL: \url{https://openreview.net/forum?id=XZm1ekzERf}.

\bibitem{DBLP:journals/tmlr/HouZLBBLJBCCKL26}
Zhaoyi~Joey Hou, Bowei~Alvin Zhang, Yining Lu, Bhiman~Kumar Baghel, Anneliese Brei, Ximing Lu, Meng Jiang, Faeze Brahman, Snigdha Chaturvedi, Haw{-}Shiuan Chang, Daniel Khashabi, and Xiang~Lorraine Li.
\newblock {CreativityPrism: {A} Cross-Domain Evaluation Framework for Large Language Model Creativity}.
\newblock {\em Transactions on Machine Learning Research}, 2026.
\newblock URL: \url{https://openreview.net/forum?id=3pfsQcEtNC}.

\bibitem{gottweis2026accelerating}
Juraj Gottweis, Wei-Hung Weng, Alexander Daryin, Tao Tu, Petar Sirkovic, Artiom Myaskovsky, Grzegorz Glowaty, Felix Weissenberger, Alessio Orlandi, Dan Popovici, et~al.
\newblock {Accelerating Scientific Discovery with Co-Scientist}.
\newblock {\em Nature}, pages 487--496, 2026.
\newblock \href {https://doi.org/10.1038/s41586-026-10644-y} {\path{doi:10.1038/s41586-026-10644-y}}.

\bibitem{DBLP:conf/iclr/KirkMNLHGR24}
Robert Kirk, Ishita Mediratta, Christoforos Nalmpantis, Jelena Luketina, Eric Hambro, Edward Grefenstette, and Roberta Raileanu.
\newblock {Understanding the Effects of {RLHF} on {LLM} Generalisation and Diversity}.
\newblock In {\em International Conference on Learning Representations ({ICLR})}, 2024.
\newblock URL: \url{https://openreview.net/forum?id=PXD3FAVHJT}.

\bibitem{2026.EDM.poster-demo-papers.424}
Manh~Hung Nguyen, Sebastian Tschiatschek, and Adish Singla.
\newblock {Enhancing Diversity of LLM-Generated Educational Tasks}.
\newblock In {\em Proceedings of the International Conference on Educational Data Mining (EDM)}, 2026.
\newblock \href {https://doi.org/10.5281/zenodo.21039728} {\path{doi:10.5281/zenodo.21039728}}.

\bibitem{hamid2026polychromic}
Jubayer Hamid, Ifdita Orney, Ellen Xu, Chelsea Finn, and Dorsa Sadigh.
\newblock {Polychromic Objectives for Reinforcement Learning}.
\newblock In {\em International Conference on Learning Representations (ICLR)}, 2026.
\newblock URL: \url{https://openreview.net/forum?id=zzTQISAGUp}.

\bibitem{finke1992creative}
Ronald~A Finke, Thomas~B Ward, and Steven~M Smith.
\newblock {\em {Creative Cognition: Theory, Research, and Applications}}.
\newblock 1992.
\newblock \href {https://doi.org/10.7551/mitpress/7722.001.0001} {\path{doi:10.7551/mitpress/7722.001.0001}}.

\bibitem{samuelson2023generative}
Pamela Samuelson.
\newblock {Generative AI meets copyright}.
\newblock {\em Science}, 381(6654):158--161, 2023.
\newblock \href {https://doi.org/10.1126/science.adi0656} {\path{doi:10.1126/science.adi0656}}.

\bibitem{DBLP:conf/aies/JiangBCKGWHFG23}
Harry~H. Jiang, Lauren Brown, Jessica Cheng, Mehtab Khan, Abhishek Gupta, Deja Workman, Alex Hanna, Johnathan Flowers, and Timnit Gebru.
\newblock {{AI} Art and its Impact on Artists}.
\newblock In {\em {AAAI/ACM} Conference on AI, Ethics, and Society {(AIES)}}, pages 363--374, 2023.
\newblock \href {https://doi.org/10.1145/3600211.3604681} {\path{doi:10.1145/3600211.3604681}}.

\bibitem{DBLP:conf/aies/LovatoZSDK24}
Juniper~L. Lovato, Julia~Witte Zimmerman, Isabelle Smith, Peter Dodds, and Jennifer~L. Karson.
\newblock {Foregrounding Artist Opinions: {A} Survey Study on Transparency, Ownership, and Fairness in {AI} Generative Art}.
\newblock In {\em Proceedings of the {AAAI/ACM} Conference on AI, Ethics, and Society {(AIES)}}, pages 905--916, 2024.
\newblock \href {https://doi.org/10.1609/AIES.V7I1.31691} {\path{doi:10.1609/AIES.V7I1.31691}}.

\bibitem{DBLP:journals/tmlr/SadasivanKB0F25}
Vinu~Sankar Sadasivan, Aounon Kumar, Sriram Balasubramanian, Wenxiao Wang, and Soheil Feizi.
\newblock {Can AI-Generated Text be Reliably Detected? Stress Testing {AI} Text Detectors Under Various Attacks}.
\newblock {\em Transactions on Machine Learning Research}, 2025.
\newblock URL: \url{https://openreview.net/forum?id=OOgsAZdFOt}.

\bibitem{DBLP:journals/patterns/LiangYMWZ23}
Weixin Liang, Mert Y{\"{u}}ksekg{\"{o}}n{\"{u}}l, Yining Mao, Eric Wu, and James Zou.
\newblock {{GPT} Detectors are Biased Against Non-native English Writers}.
\newblock {\em Patterns}, 4(7):100779, 2023.
\newblock \href {https://doi.org/10.1016/J.PATTER.2023.100779} {\path{doi:10.1016/J.PATTER.2023.100779}}.

\bibitem{DBLP:conf/chi/HeHW25}
Jessica He, Stephanie Houde, and Justin~D. Weisz.
\newblock {Which Contributions Deserve Credit? Perceptions of Attribution in Human-AI Co-Creation}.
\newblock In {\em Proceedings of the {CHI} Conference on Human Factors in Computing Systems ({CHI})}, pages 540:1--540:18, 2025.
\newblock \href {https://doi.org/10.1145/3706598.3713522} {\path{doi:10.1145/3706598.3713522}}.

\bibitem{DBLP:conf/icml/KirchenbauerGWK23}
John Kirchenbauer, Jonas Geiping, Yuxin Wen, Jonathan Katz, Ian Miers, and Tom Goldstein.
\newblock {A Watermark for Large Language Models}.
\newblock In {\em International Conference on Machine Learning ({ICML})}, pages 17061--17084, 2023.
\newblock URL: \url{https://proceedings.mlr.press/v202/kirchenbauer23a.html}.

\bibitem{DBLP:journals/nature/DathathriSGHMWBKSMHVMBB24}
Sumanth Dathathri, Abigail See, Sumedh Ghaisas, Po{-}Sen Huang, Rob McAdam, Johannes Welbl, Vandana Bachani, Alex Kaskasoli, Robert Stanforth, Tatiana Matejovicova, Jamie Hayes, Nidhi Vyas, Majd~Al Merey, Jonah Brown{-}Cohen, Rudy Bunel, Borja Balle, A.~Taylan Cemgil, Zahra Ahmed, Kitty Stacpoole, Ilia Shumailov, Ciprian Baetu, Sven Gowal, Demis Hassabis, and Pushmeet Kohli.
\newblock {Scalable Watermarking for Identifying Large Language Model Outputs}.
\newblock {\em Nature}, 634(8035):818--823, 2024.
\newblock \href {https://doi.org/10.1038/S41586-024-08025-4} {\path{doi:10.1038/S41586-024-08025-4}}.

\bibitem{DBLP:conf/imc/0001LSVZS25}
Enze Liu, Elisa Luo, Shawn Shan, Geoffrey~M. Voelker, Ben~Y. Zhao, and Stefan Savage.
\newblock {Somesite {I} Used To Crawl: Awareness, Agency and Efficacy in Protecting Content Creators From {AI} Crawlers}.
\newblock In {\em Proceedings of the {ACM} Internet Measurement Conference ({IMC})}, pages 78--99, 2025.
\newblock \href {https://doi.org/10.1145/3730567.3732913} {\path{doi:10.1145/3730567.3732913}}.

\bibitem{DBLP:journals/bjet/FanTLSTZSLG25}
Yizhou Fan, Luzhen Tang, Huixiao Le, Kejie Shen, Shufang Tan, Yueying Zhao, Yuan Shen, Xinyu Li, and Dragan Gasevic.
\newblock {Beware of Metacognitive Laziness: Effects of Generative Artificial Intelligence on Learning Motivation, Processes, and Performance}.
\newblock {\em British Journal of Educational Technology}, 56(2):489--530, 2025.
\newblock \href {https://doi.org/10.1111/BJET.13544} {\path{doi:10.1111/BJET.13544}}.

\bibitem{DBLP:conf/chi/LeeSTDRBW25}
Hao{-}Ping~(Hank) Lee, Advait Sarkar, Lev Tankelevitch, Ian Drosos, Sean Rintel, Richard Banks, and Nicholas~C. Wilson.
\newblock {The Impact of Generative {AI} on Critical Thinking: Self-Reported Reductions in Cognitive Effort and Confidence Effects From a Survey of Knowledge Workers}.
\newblock In {\em Proceedings of the {CHI} Conference on Human Factors in Computing Systems ({CHI})}, pages 1121:1--1121:22, 2025.
\newblock \href {https://doi.org/10.1145/3706598.3713778} {\path{doi:10.1145/3706598.3713778}}.

\bibitem{DBLP:conf/chi/SuhCMLX24}
Sangho Suh, Meng Chen, Bryan Min, Toby~Jia{-}Jun Li, and Haijun Xia.
\newblock {Luminate: Structured Generation and Exploration of Design Space with Large Language Models for Human-AI Co-Creation}.
\newblock In {\em Proceedings of the {CHI} Conference on Human Factors in Computing Systems ({CHI})}, pages 644:1--644:26, 2024.
\newblock \href {https://doi.org/10.1145/3613904.3642400} {\path{doi:10.1145/3613904.3642400}}.

\bibitem{DBLP:journals/pacmhci/BucincaMG21}
Zana Bu{\c{c}}inca, Maja~Barbara Malaya, and Krzysztof~Z. Gajos.
\newblock {To Trust or to Think: Cognitive Forcing Functions Can Reduce Overreliance on {AI} in AI-assisted Decision-making}.
\newblock {\em ACM on Human-Computer Interaction}, 5({CSCW1}):188:1--188:21, 2021.
\newblock \href {https://doi.org/10.1145/3449287} {\path{doi:10.1145/3449287}}.

\bibitem{DBLP:journals/pacmhci/VasconcelosJGGBK23}
Helena Vasconcelos, Matthew J{\"{o}}rke, Madeleine Grunde{-}McLaughlin, Tobias Gerstenberg, Michael~S. Bernstein, and Ranjay Krishna.
\newblock {Explanations Can Reduce Overreliance on {AI} Systems During Decision-Making}.
\newblock {\em ACM on Human-Computer Interaction}, 7({CSCW1}):1--38, 2023.
\newblock \href {https://doi.org/10.1145/3579605} {\path{doi:10.1145/3579605}}.

\bibitem{DBLP:journals/corr/abs-2512-04630}
Heeryung Choi, Tung Phung, Mengyan Wu, Adish Singla, and Christopher Brooks.
\newblock {Reflection-Satisfaction Tradeoff: Investigating Impact of Reflection on Student Engagement with AI-Generated Programming Hints}.
\newblock {\em CoRR}, abs/2512.04630, 2025.
\newblock \href {https://doi.org/10.48550/ARXIV.2512.04630} {\path{doi:10.48550/ARXIV.2512.04630}}.

\bibitem{doi:10.1126/sciadv.adn5290}
Anil~R. Doshi and Oliver~P. Hauser.
\newblock {Generative AI enhances individual creativity but reduces the collective diversity of novel content}.
\newblock {\em Science Advances}, 10(28):eadn5290, 2024.
\newblock \href {https://doi.org/10.1126/sciadv.adn5290} {\path{doi:10.1126/sciadv.adn5290}}.

\end{thebibliography}
\bibliographystyle{unsrturl}

\end{document}